\documentstyle[a4,11pt]{article}
\begin{document}

\begin{titlepage}
\vskip 2cm
\begin{flushright}
Preprint CNLP-1997-02
\end{flushright}
\vskip 2cm
\begin{center}
{\large {\bf SOLITONS, SURFACES, CURVES, AND THE SPIN DESCRIPTION
OF NONLINEAR EVOLUTION EQUATIONS}}\footnote{Preprint
CNLP-1997-02.Alma-Ata.1997 \\
cnlpmyra@satsun.sci.kz}
\vskip 2cm

{\bf R. Myrzakulov$^{a,b}$  }

\end{center}
\vskip 1cm
$^{a}$Physical-Thechnical Institute, NAS RK, 480082, Alma-Ata-82, Kazakhstan \\
$^{b}$Centre for Nonlinear Problems, PO Box 30, 480035, Alma-Ata-35, Kazakhstan

\vskip 1cm

\begin{abstract}
The briefly review on the common spin description of the nonlinear
evolution equations.
\end{abstract}


\end{titlepage}

\setcounter{page}{1}
\newpage

\begin{center}
{\bf I.INTRODUCTION}
\end{center}

In 1977 Lakshmanan[1] discovered that between the isotropic Landau-Lifshitz equation(LLE)
$$ \vec S_{t}=\vec S\wedge \vec S_{xx} \eqno(1)$$
and the NLSE
$$ iq_t+q_{xx}+2q^{2}\bar q = 0 \eqno(2)$$
takes place the equivalence, which following to the terminology of ref[2]
we call the Lakshmanan equivalence(shortly, L-equivalence).
On the other hand, as well known equations(1) and (2) are gauge equivalent
to each other[3]. This equivalence we call gauge equivalence or briefly
g-equivalence. In a series of papers [2] were presented the new class
integrable and nonintegrale spin equations
and the unit spin description of soliton equations.  In this paper using
the geometrical method we will present the Lakshmanan equivalent
counterparts of the some spin equations.

Consider a n-dimensional space with the basic unit vectors: $\vec e_{1}=
\vec S, \vec e_{2}, ... ,\vec e_{n}$. Then the M-0 equation
has the form[2]
$$
\vec S_{t} = \sum^{n}_{i=2} a_{i} \vec e_{i}          \eqno(3)
$$
where $a_{i}$ are real functions, $\vec S = (S_{1}, S_{2}, ... , S_{n}),
\vec S^{2} = E = \pm 1$. This equation admits the many interesting class
integrable and nonintegrable reductions[2]. Below we present the
L-equivalent counterparts of the some  reductions and
only for the cases $n = 2, 3$.

The paper is organized as follows. In Sec.II and Sec.III
we briefly review some necessary
informations about the some surfaces/curves approaches to the nonlinear evolution equations.
The L-equivalent counterparts of the some spin equations are presented
in Sec.VI and Sec.V. Integrals of motion are discussed in Sec.VI.
In Sec.VII the Lax representations are derived. A non-isospectral problems are
discussed in Sec.VIII. The solitons (line and curved), domain walls
and dromion-like solutions of the M-I equation as well as their breaking
analogues are obtained in
Sec.IX and breaking solitons and dromions of the Zakharov equation  are
obtained in Sec.X. In Sec.XI we present the L-equivalent
counterparts of the some spin-phonon systems
and we conclude in Sec.XII.
\\
\\
\\
\begin{center}
{\bf II.THE D - APPROACH}
\end{center}
\begin{center}
{\bf IIa.The n-dimensional case}
\end{center}

Consider a space $R_{n}$ with an orthonormal basis
$\vec \xi_{i},  i=1,2,...,n$.
In the D-approach the fundamental set of equations for example in 2+1 dimensions
looks like[2]
$$ \vec r_{t} =  \sum^{n}_{i=1} b_{i} \vec \xi_{i}          \eqno(4a)$$
$$ \vec \xi_{1x} =  k_{1} \vec \xi_{2}          \eqno(4b)$$
$$ \vec \xi_{jx} =  Ek_{j} \vec \xi_{j-1} + k_{j+1}\vec \xi_{j+1}  \eqno(4c)$$
$$ \vec \xi_{nx} =  Ek_{n} \vec \xi_{n-1}          \eqno(4d)$$
$$ \vec \xi_{iy} =  \sum^{n}_{m=1} a_{im}\vec \xi_{m} \eqno(4e)$$
$$ \vec \xi_{it} =  \sum^{n}_{m=1} b_{im}\vec \xi_{m} \eqno(4f)$$
where $j=2,3,...,n-1; a_{ii}=b_{ii}=0 $. Note that eqs(4b)-(4d) are
the Serret-Frenet equations(SFE).
\begin{center}
{\bf IIb.The (2+1)-dimensions: a space curves}
\end{center}
In this case equations of the curves have the forms[2]
$$
\left ( \begin{array}{ccc}
\vec e_{1} \\
\vec e_{2} \\
\vec e_{3}
\end{array} \right)_{x} = C
\left ( \begin{array}{ccc}
\vec e_{1} \\
\vec e_{2} \\
\vec e_{3}
\end{array} \right)  \eqno(5a)
$$

$$
\left ( \begin{array}{ccc}
\vec e_{1} \\
\vec e_{2} \\
\vec e_{3}
\end{array} \right)_{y} = D
\left ( \begin{array}{ccc}
\vec e_{1} \\
\vec e_{2} \\
\vec e_{3}
\end{array} \right) \eqno(5b)
$$

Here
\begin{displaymath}
C =
\left ( \begin{array}{ccc}
0   & k     & 0 \\
-Ek & 0     & \tau  \\
0   & -\tau & 0
\end{array} \right) ,
\end{displaymath}
\begin{displaymath}
D =
\left ( \begin{array}{ccc}
0       & m_{3}  & -m_{2} \\
-Em_{3} & 0      & m_{1} \\
Em_{2}  & -m_{1} & 0
\end{array} \right)
\end{displaymath}
Hence now we have
$$C_y - D_x + [C, D] = 0 \eqno (6) $$

The time evolution of $\vec e_{i}$ we can write in the form[1]
$$
\left ( \begin{array}{ccc}
\vec e_{1} \\
\vec e_{2} \\
\vec e_{3}
\end{array} \right)_{t} = G
\left ( \begin{array}{ccc}
\vec e_{1} \\
\vec e_{2} \\
\vec e_{3}
\end{array} \right) \eqno(7)
$$
with
\begin{displaymath}
G =
\left ( \begin{array}{ccc}
0       & \omega_{3}  & -\omega_{2} \\
-E\omega_{3} & 0      & \omega_{1} \\
E\omega_{2}  & -\omega_{1} & 0
\end{array} \right) ,
\end{displaymath}
So, we have
$$C_t - G_x + [C, G] = 0 \eqno (8a) $$
$$D_t - G_y + [D, G] = 0 \eqno (8b) $$
\\
\begin{center}
{\bf IIc.The (2+1)-dimensions: a plane curves}
\end{center}
In this case we have[2]
$$
\left ( \begin{array}{ccc}
\vec e_{1} \\
\vec e_{2}
\end{array} \right)_{x} = C
\left ( \begin{array}{ccc}
\vec e_{1} \\
\vec e_{2}
\end{array} \right)  \eqno(9a)
$$

$$
\left ( \begin{array}{ccc}
\vec e_{1} \\
\vec e_{2}
\end{array} \right)_{y} = D
\left ( \begin{array}{ccc}
\vec e_{1} \\
\vec e_{2}
\end{array} \right) \eqno(9b)
$$
Here
\begin{displaymath}
C =
\left ( \begin{array}{ccc}
0   & k     \\
-Ek & 0
\end{array} \right) ,
\end{displaymath}
\begin{displaymath}
D =
\left ( \begin{array}{ccc}
0       & m_{3} \\
-Em_{3} & 0
\end{array} \right)
\end{displaymath}
Now from the MPCE (6)  we obtain
$$  m_{3} =
\partial ^{-1}_{x}k_{y} \eqno(10) $$
For the time evolution we get[1]
$$
\left ( \begin{array}{ccc}
\vec e_{1} \\
\vec e_{2}
\end{array} \right)_{t} = G
\left ( \begin{array}{ccc}
\vec e_{1} \\
\vec e_{2}
\end{array} \right) \eqno(11)
$$
where
$$
G =
\left ( \begin{array}{ccc}
0       & \omega_{3}  \\
-E\omega_{3} & 0
\end{array} \right).
$$

Now already we ready using this formalism  to construct the L-equivalent
counterparts of the some spin equations.
\\
\\
\\
\begin{center}
{\bf III.THE C - APPROACH}
\end{center}
\begin{center}
{\bf IIIa.A hypersurface $V_{n-1}$ in $R_{n}$}
\end{center}

In this case we consider a time-dependent (n-1)-dimensional hypersurface
$V_{n-1}(t,\vec r) $ in $R_{n}$[2]. Define
$$
\vec \xi_{j} = \vec r_{x^{j}},  \,\,\,\,j=1,2,...,n-1   \eqno(12)
$$
and $\vec \nu $ is the unit normal vector to the hypersurface. The
first and second fundamental forms  are given by
$$ I=d\vec r d\vec r = \sum^{n-1}_{i,j=1}g_{ij}dx^{i}dx^{j} \eqno (13a) $$
$$ II=-d\vec r d\vec \nu = \sum^{n-1}_{i,j=1}b_{ij}dx^{i}dx^{j} \eqno (13b) $$

In the C-approach[2], the fundamental set of equations reads as
$$ \vec r_{t} =  \sum^{n-1}_{i=1} b_{i} \vec \xi_{i} +b_{0}\vec \nu \eqno(14a)  $$
$$ \bigtriangledown_{i}\vec \xi_{j} = b_{ij}\vec \nu    \eqno(14b)$$
$$ \bigtriangledown_{i}\vec \nu = -E b^{j}_{i}\vec \xi_{j}    \eqno(14c)$$
here that eqs(14b,c) are the Gauss-Weingarten equations. Then the Peterson-Mainardi-Codazzi equation (MPCE) takes the form
$$ R_{ij,kl} = E(b_{ik}b_{jl}-b_{jk}b_{il})    \eqno(15a)$$
$$ \bigtriangledown_{i}b_{jk}  = \bigtriangledown_{j}b_{ik}    \eqno(15b)$$
where $ \bigtriangledown_{i}$ is the covariant derivative.

\begin{center}
{\bf IIIb. The 2+1 dimensional case}
\end{center}

In this case, starting from the
results of the ref.[4](see, also, for example, [5-7]) we consider the motion of surface in
the 3-dimensional space which generated by a position vector
$\vec r(x,y,t) = \vec r(x^{1}, x^{2}, t)$. According to the C-approach[2],
the main elements of which we present in this section,
 let x and y are local coordinates on the surface. The
first and second fundamental forms in the usual notation are given by
$$ I=d\vec r d\vec r=Edx^2+2Fdxdy+Gdy^2 \eqno (16a) $$
$$ II=-d\vec r d\vec n=Ldx^2+2Mdxdy+Ndy^2 \eqno (16b) $$
where $$ E=\vec r_x\vec r_x=g_{11}, F=\vec r_x\vec r_y=g_{12},
G=\vec r_{y^2}=g_{22},   $$
$$ L=\vec n\vec r_{xx}=b_{11}, M=\vec n\vec r_{xy}=b_{12},
N=\vec n\vec r_{yy}=b_{22}, \vec n=\frac{(\vec r_x \wedge \vec r_y)}
{|\vec r_x \wedge \vec r_y|}  $$

In this case, the starting set of equations of the C-approach[2],
becomes
$$ \vec r_{t} = a \vec r_x + b \vec r_y + c \vec n \eqno (17a) $$
$$ \vec r_{xx}=\Gamma^1_{11} \vec r_x + \Gamma^2_{11} \vec r_y +L \vec n \eqno (17b) $$
$$ \vec r_{xy}=\Gamma^1_{12} \vec r_x + \Gamma^2_{12} \vec r_y +M \vec n \eqno (17c)$$
$$ \vec r_{yy}=\Gamma^1_{22} \vec r_x + \Gamma^2_{22} \vec r_y + N \vec n \eqno (17d)$$
$$ \vec n_x=p_1 \vec r_x+p_2 \vec r_y \eqno (17e) $$
$$ \vec n_y=q_1 \vec r_x+q_2 \vec r_y \eqno (17f) $$
where $ \Gamma^k_{ij} $ are the Christoffel symbols of the second kind defined by
the metric $ g_{ij} $ and $ g^{ij}=(g_{ij})^{-1} $ as
$$ \Gamma^k_{ij}=\frac{1}{2} g^{kl}(\frac{\partial g_{lj}} {\partial x^i}+
   \frac {\partial g_{il}}{ \partial x^j}-\frac{\partial g_{ij}}
   {\partial x^l}) \eqno (18) $$
The coefficients $ p_i, q_i $ are given by
$$ p_i=-b_{1j}g^{ji}, \,\,\, q_i=-b_{2j}g^{ji} \eqno (19) $$
The compatibility conditions $ \vec r_{xxy}=\vec r_{xyx} $ and
$ \vec r_{yyx}=\vec r_{xyy} $ yield the following Mainardi-Peterson-Codazzi
equations (MPCE)
$$ R^l_{ijk} = b_{ij}b^l_{k}-b_{ik}b^l_{j} \eqno (20a) $$
$$ \frac{\partial b_{ij}}{\partial x^k}-\frac{\partial b_{ik}}{\partial
    x^j}=\Gamma^s_{ik}b_{is}-\Gamma^s_{ij}b_{ks}
\eqno (20b) $$
where $ b^j_i=g^{jl}b_{il} $ and the curvature tenzor has the form
$$ R^l_{ijk} = \frac{\partial \Gamma^l_{ij}}{\partial x^k}-\frac{\partial \Gamma^l_{ik}}
{\partial x^j}+\Gamma^s_{ij} \Gamma^l_{ks}-\Gamma^s_{ik} \Gamma^l_{js}
\eqno (21) $$

Let
$\vec Z = ( r_{x},  r_{y},  n)^{t}$ . Then
$$ \vec Z_{x} = A \vec Z \eqno (22a) $$
$$ \vec Z_{y} = B \vec Z \eqno (22b) $$
where
\begin{displaymath}
A =
\left ( \begin{array}{ccc}
\Gamma^{1}_{11} & \Gamma^{2}_{11} & L \\
\Gamma^{1}_{12} & \Gamma^{2}_{12} & M \\
p_{1}           & p_{2}           & 0
\end{array} \right) ,
\end{displaymath}
\begin{displaymath}
B =
\left ( \begin{array}{ccc}
\Gamma^{1}_{12} & \Gamma^{2}_{12} & M \\
\Gamma^{1}_{22} & \Gamma^{2}_{22} & N \\
q_{1}           & q_{2}           & 0
\end{array} \right)
\end{displaymath}
Hence we get the new form of the MPCE(20)
$$ A_y - B_x + [A, B] = 0 \eqno (23) $$

Let us introduce the orthogonal trihedral[1]
$$ \vec e_{1} = \frac{\vec r_x}{\surd E}, \,\,\,
\vec e_{2} = \frac{\vec r_y}{\surd G}, \,\,\, \vec e_{3} = \vec e_{1} \wedge
\vec e_{2} = \vec n    \eqno(24a) $$
or
$$ \vec e_{1} = \frac{\vec r_x}{\surd E}, \,\,\,
\vec e_{2} = \vec n, \,\,\, \vec e_{3} = \vec e_{1} \wedge
\vec e_{2}    \eqno(24b) $$

Let $ \vec r_x^2 = E = \pm 1$.  Then from the previouse equations after some algebra
we get[2]
$$
\left ( \begin{array}{ccc}
\vec e_{1} \\
\vec e_{2} \\
\vec e_{3}
\end{array} \right)_{x}= C
\left ( \begin{array}{ccc}
\vec e_{1} \\
\vec e_{2} \\
\vec e_{3}
\end{array} \right)  \eqno(25a)
$$

$$
\left ( \begin{array}{ccc}
\vec e_{1} \\
\vec e_{2} \\
\vec e_{3}
\end{array} \right)_{y} = D
\left ( \begin{array}{ccc}
\vec e_{1} \\
\vec e_{2} \\
\vec e_{3}
\end{array} \right) \eqno(25b)
$$

Here
\begin{displaymath}
C =
\left ( \begin{array}{ccc}
0   & k     & -\sigma \\
-Ek & 0     & \tau  \\
E\sigma   & -\tau & 0
\end{array} \right) ,
\end{displaymath}
\begin{displaymath}
D =
\left ( \begin{array}{ccc}
0       & m_{3}  & -m_{2} \\
-Em_{3} & 0      & m_{1} \\
Em_{2}  & -m_{1} & 0
\end{array} \right)
\end{displaymath}
Now the MPCE (20) and/or (23) becomes
$$C_y - D_x + [C, D] = 0 \eqno (26) $$
Hence as $\sigma = 0$ we obtain
$$ (m_{1}, m_{2}, m_{3}) =
(\partial ^{-1}_{x}(\tau_{y} + k m_{2}), m_{2},
\partial ^{-1}_{x}(k_{y} - \tau m_{2}) ) \eqno(27a) $$
$$ m_{2} = \partial ^{-1}_{x}(\tau m_{3} - k m_{1}) \eqno(27b) $$

The time evolution of $\vec e_{i}$ we can write in the form[2]
$$
\left ( \begin{array}{ccc}
\vec e_{1} \\
\vec e_{2} \\
\vec e_{3}
\end{array} \right)_{t} = G
\left ( \begin{array}{ccc}
\vec e_{1} \\
\vec e_{2} \\
\vec e_{3}
\end{array} \right) \eqno(28)
$$
with
\begin{displaymath}
G =
\left ( \begin{array}{ccc}
0       & \omega_{3}  & -\omega_{2} \\
-E\omega_{3} & 0      & \omega_{1} \\
E\omega_{2}  & -\omega_{1} & 0
\end{array} \right) ,
\end{displaymath}
So, we have
$$C_t - G_x + [C, G] = 0 \eqno (29a) $$
$$D_t - G_y + [D, G] = 0 \eqno (29b) $$
Below using these C-, and D-approches we will construct
the L-equivalent counterparts of the some spin systems.
\\
\\
\\
\begin{center}
{\bf IV. THE LAKSHMANAN EQUIVALENT COUNTERPARTS OF THE
SOME MYRZAKULOV EQUATIONS: the 2-dimensional case}
\end{center}

In this case the Myrzakulov-0 equation(3) becomes
$$
\vec S_{t} = a_{2} \vec e_{2}          \eqno(30)
$$
and  $\vec S = (S_{1}, S_{2}),
\vec S^{2} = E = \pm 1, \tau = c =0$.
So, we have[2]
$$
\left ( \begin{array}{ccc}
\vec e_{1} \\
\vec e_{2}
\end{array} \right)_{x} = C
\left ( \begin{array}{ccc}
\vec e_{1} \\
\vec e_{2}
\end{array} \right)  \eqno(31a)
$$

$$
\left ( \begin{array}{ccc}
\vec e_{1} \\
\vec e_{2}
\end{array} \right)_{y} = D
\left ( \begin{array}{ccc}
\vec e_{1} \\
\vec e_{2}
\end{array} \right) \eqno(31b)
$$
Here
\begin{displaymath}
C =
\left ( \begin{array}{ccc}
0   & k     \\
-Ek & 0
\end{array} \right) ,
\end{displaymath}
\begin{displaymath}
D =
\left ( \begin{array}{ccc}
0       & m_{3} \\
-Em_{3} & 0
\end{array} \right)
\end{displaymath}
Now from the MPCE (20)  we obtain
$$  m_{3} =
\partial ^{-1}_{x}k_{y} \eqno(32) $$
For the time evolution we get[1]
$$
\left ( \begin{array}{ccc}
\vec e_{1} \\
\vec e_{2}
\end{array} \right)_{t} = G
\left ( \begin{array}{ccc}
\vec e_{1} \\
\vec e_{2}
\end{array} \right) \eqno(33)
$$
where
$$
G =
\left ( \begin{array}{ccc}
0       & \omega_{3}  \\
-E\omega_{3} & 0
\end{array} \right).
$$

Now already we ready using this formalism  to construct the L-equivalent
counterparts of the some Myrzakulov equations.
\\
\\
{\bf Examples}
\\

1) The Myrzakulov-IV(M-IV) equation[2]
$$ \vec S_{t}+\{\vec S_{xy}+V\vec S +
E\vec S_{x}\wedge (\vec S\wedge\vec S_{y})\}_{x}=0 \eqno(34a) $$
$$V_{x}=\frac{E}{2} (\vec S^{2}_{x})_{y}    \eqno(34b) $$

In this case
$$ \vec r_{t} = W \vec e_{1} + U \vec e_{2}  \eqno (35a) $$
or
$$ \vec r_{t} =-V \vec e_{1} - k_{y} \vec e_{2}  \eqno (35b) $$
So, we have
$$ W = -V,\,\,\,\,\, U=-k_{y}    \eqno(36a)$$
$$ W_{x} = EkU    \eqno(36b)                 $$
$$ k_{t} = U_{xx}+Ek_{x}\partial^{-1}_{x}(kU)+Ek^{2}U    \eqno(37b) $$
Hence we get the following L-equivalent counterpart of the M-IV equation(34)
$$ k_{t} +k_{xxy}+(kV)_{x}=0    \eqno(38a) $$
$$ V_{x}=Ekk_{y}    \eqno(38b) $$
which is the 2+1 dimensional mKdV[10]. Note that in the our case $E=\pm 1$.

Similarly we can find the L-equivalent counterparts of the some other
Myrzakulov equations. Now we present the final results. Details of these
calculations are given in the our ealier papers(see, for example, [8-9]).

2) The M-XXI equation has the following L-equivalent counterpart
$$ k_{t} +k_{xxy}-2(kV_{1})_{x}+2V_{2}k=0    \eqno(39a) $$
$$ V_{1x}=Ek_{y}    \eqno(39b) $$
$$ V_{2x}=-Ek_{xy}    \eqno(39c) $$
which is the 2+1 dimensional KdV[10]. Note that in the our case $E=\pm 1$.

Now after the some minor modifications of the above presented
formalism we obtain the following L-equivalents
of the some spin systems.\\
3) The M-XXIII equation   has the following L-equivalent
$$
k_{xx}-\sigma^2k_{yy}+\frac{1}{2}R(x,y,t)e^{2k}+(2k_{tt}+3k^2_t)e^{2k}=0
\eqno (40)
$$
4) The M-XXIV equation   has the following L-equivalent
$$
k_{xx}+\sigma^2k_{yy}+\frac{1}{4}R\sin k+\sin k(k_{tt}-\frac{1}{2}k_t^2)=0
\eqno (41)
$$
5) The M-XXV equation   has the following L-equivalent
$$
(\frac{k_{tx}}{\cos k})_x\sin k-\sigma^2(\frac{k_{ty}}{\sin k}_y\cos k +
\frac{1}{2}(k_x^2)_t+ \frac{1}{2}(k_y^2)_t+(k_{xx}+ \sigma^2k_{yy})k_t \\
=(\frac{1}{2}R+3)\sin k\cos kk_t. \eqno (42)
$$
6) The M-XXVI equation   has the following L-equivalent
$$
(k_{xx} +k_{yy})_t +(k_{xx} +k_{yy})k_t=(- \frac {1}{2}R -3)e^{2k}k_t
\eqno (43)
$$
7) The M-XXVII equation   has the following L-equivalent
$$
k_{xx} +k_{yy} - A(x,y)e^{-k} +(1+ \frac {R(x,y)}{3})e^{2k} =0
\eqno(44)
$$
8)The M-XXVIII equation   has the following L-equivalent
$$ k_{xx} +k_{yy} +k_{tt} - \frac{R}{8} k^5 =0 \eqno (45)$$
Of course we can get these results using the other approaches(see, for example,
[8-9]). Note that eqs(40)-(45) were considered in [11] from the other
point of view.
\\
\\
\\
\begin{center}
{\bf V. THE LAKSHMANAN EQUIVALENT COUNTERPARTS OF THE
ISHIMORI and SOME MYRZAKULOV EQUATIONS: the 3-dimensional case}
\end{center}

In this case work equations(25)-(29) and the Myrzakulov-0 equation becomes
$$
\vec S_{t} = a_{2} \vec e_{2} +a_{3}\vec e_{3}   \eqno (46)
$$
and  $\vec S = (S_{1}, S_{2}, S_{3}),
\vec S^{2} = E = \pm 1$. Using these equations we construct
the L-equivalent counterparts of the some Myrzakulov and Ishimori
equations[2].  Below we present  only the final results.
\\
{\bf Examples}
\\

1) The Myrzakulov-I(M-I) equation looks like[2]
$$ \vec S_{t}=(\vec S\wedge \vec S_{y}+u\vec S)_x \eqno (47a) $$
$$ u_x=-\vec S_(\vec S_{x}\wedge \vec S_{y}) \eqno (47b) $$
In this case the fundamental set of the equations of the  C-approach[2]
becomes
$$ \vec r_{t} = u \vec r_x - M(\frac{E}{G})^{1/2} \vec r_y +
\frac{G_{x}}{2}(\frac{E}{G})^{1/2}m_{2} \vec n \eqno (48a) $$
$$ \vec r_{xx}=L \vec n \eqno (48b) $$
$$ \vec r_{xy}=\frac{G_{x}}{2G}\vec r_y +M \vec n \eqno (48c)$$
$$ \vec r_{yy}=-\frac{G_{x}}{2E}\vec r_x + \frac{G_{y}}{2G} \vec r_y + N \vec n \eqno (48d)$$
$$ \vec n_x=-EL \vec r_x-\frac{M}{G}\vec r_y \eqno (48e) $$
$$ \vec n_y=-EM \vec r_x - \frac{N}{G} \vec r_y \eqno (48f) $$
So the MPCE(23) take the form
$$ K=\frac{LN-M^{2}}{G} = \frac{G_{x}^{2}}{4G^{2}}-\frac{G_{xx}}{2G}\eqno (49a) $$
$$ L_{y}-M_{x} = M \frac{G_{x}}{2G}\eqno (49b) $$
$$ M_{y}-N_{x} = - \frac{EG_{x}}{2}L + \frac{G_{y}}{2G}M -
\frac{G_{x}}{2G}N \eqno (49c) $$
Now let us pass to the new variable
$$ q = \frac{L}{2} \exp{(-i\partial^{-1}_{x} MG^{-1/2})} \eqno(50) $$
Then this variable satisfies the following Zakharov equation
$$ iq_t=q_{xy}+2d^2Vq, \eqno(51a) $$
$$ V_x=2E(|q|^2)_y. \eqno (51b) $$
which is the L-equivalent (and the g-equivalent[12]) counterpart
of the M-I equation(47).  Similarly we can find the L-equivalents of the
some other spin equations. Details are given for example
in the refs[8-9].

2) The M-II equation[2]
$$ \vec S_{t}=(\vec S\wedge \vec S_{y}+u\vec S)_x+2cb^{2}\vec S_{y}
     -4cv\vec S_{x} \eqno (52a) $$
$$ u_x=-\vec S_(\vec S_{x}\wedge \vec S_{y}), \eqno(52b) $$
$$ v_x=\frac{1}{16b^{2}c^{2}}
   (\vec S^2_{1x})_y \eqno (52c) $$
has the following L-equivalent
$$ iq_t=q_{xy}-4ic(Vq)_x, \eqno(53a) $$
$$ V_x=2E(|q|^2)_y. \eqno (53b) $$
which is theStrachan equation[10].

3) The M-III equation[2]
$$ \vec S_{t}=(\vec S\wedge \vec S_{y}+u\vec S)_x+2b(cb+d)\vec S_{y}
     -4cv\vec S_{x} \eqno (54a) $$
$$ u_x=-\vec S_(\vec S_{x}\wedge \vec S_{y}), \eqno(54b)$$
$$ v_x=\frac{1}{4(2bc+d)^2}
   (\vec S^2_{1x})_y \eqno (54c) $$
in this case
$$ (m_1, m_2, m_3)=(\partial^{-1}_x(\tau_y+km_2), -\frac{-u_x}{k},
    \partial^{-1}_x(k_y-\tau m_2))  \eqno (55) $$
and the L-equivalent is the following set of equations [2]
$$ iq_t=q_{xy}-4ic(Vq)_x+2d^2Vq, \eqno(56a)$$
$$ V_x=2E(|q|^2)_y. \eqno (56b) $$

Note that equations (54) and (56) admit the following integrable
reductions: a) the M-I[2] and the
Zakharov[14] equations, as c=0; b) the M-II[2] and Strachan[10] equations, as d=0,
respectively [2]. Note that between the M-I,M-II,M-III  equations
and eqs(51),(53),(56) take place the g-equivalence, respectively [12].

4) The M-VIII equation looks like[2]
$$ iS_t=\frac{1}{2}[S_{xx},S]+iu_xS_x   \eqno (57a) $$
$$ u_{xy}=\frac{1}{4i}tr(S[S_y,S_x])  \eqno (57b) $$
where the subscripts denote partial derivatives and S denotes the spin
matrix $ (r^2=\pm1)$

$$S= \pmatrix{
S_3 & rS^- \cr
rS^+ & -S_3
}, \eqno (58)$$
$$ S^2=I    $$

Equations (57) are integrable, i.e. admits Lax representation
and different type soliton solutions [2]. The Lakshmanan equivalent
counterpart of the M-VIII equation (57) has the form[2]
$$ iq_t + q_{xx} + v q = 0,  \eqno(59a) $$
$$ ip_t - p_{xx} - v p = 0,\eqno(59b) $$
$$ v_y=2(pq)_x \eqno(59c) $$
where $p = E\bar q$. On the other hand, in [12] was shown that eqs.(57)
and (59) are gauge equivalent to each other.

5) The Ishimori equation
$$ iS_t+\frac{1}{2}[S,M_{10}S]+A_{20}S_x+A_{10}S_y = 0 \eqno(60a)$$
$$ M_{20}u=\frac{\alpha}{4i}tr(S[S_y,S_x]) \eqno(60b)$$
where $ \alpha,b,a  $= consts and
$$ M_{j0} = M_{j},\,\,\, A_{j0}=A_{j}\,\,\,\, as \, \,\,\,a = b = -\frac{1}{2}. $$

The L-equivalent counterpart has the form [1]
$$ iq_t+M_{10}q+vq=0 \eqno(61a)$$
$$ ip_t-M_{10}p-vp=0 \eqno(61b)$$
$$ M_{20}v=M_{10}(pq) \eqno(61c)$$
which is the Davey-Stewartson equation, where $p=E\bar q$.
As well known these equations are too gauge equivalent to each other[13].

6) The M-IX equation has the form[2]
$$ iS_t+\frac{1}{2}[S,M_1S]+A_2S_x+A_1S_y = 0 \eqno(62a)$$
$$ M_2u=\frac{\alpha}{4i}tr(S[S_y,S_x]) \eqno(62b)$$
where $ \alpha,b,a  $=  consts and
$$ M_1= \alpha ^2\frac{\partial ^2}{\partial y^2}-2\alpha (b-a)\frac{\partial^2}
   {\partial x \partial y}+(a^2-2ab-b)\frac{\partial^2}{\partial x^2}; $$
$$ M_2=\alpha^2\frac{\partial^2}{\partial y^2} -\alpha(2a+1)\frac{\partial^2}
   {\partial x \partial y}+a(a+1)\frac{\partial^2}{\partial x^2},$$
$$ A_1=i\alpha\{(2ab+a+b)u_x-(2b+1)\alpha u_y\} $$
$$ A_2=i\{\alpha(2ab+a+b)u_y-(2a^2b+a^2+2ab+b)u_x\}, $$
Eqs.(62) admit the two integrable reductions. As b=0, eqs. (62)
after the some manipulations reduces to the M-VIII equation (57) and as
 $ a=b=-\frac{1}{2} $
to the Ishimori equation(60). In general we have the two integrable cases:
 the M-IXA equation as $\alpha^{2} = 1,$ the M-IXB equation
as $\alpha^{2} = -1$. We note that the M-IX equation is integrable and
admits the following Lax representation [2]
$$ \alpha \Phi_y =\frac{1}{2}[S+(2a+1)I]\Phi_x \eqno(63a) $$
$$ \Phi_t=\frac{i}{2}[S+(2b+1)I]\Phi_{xx}+\frac{i}{2}W\Phi_x \eqno(63b) $$
where $$ W_1=W-W_2=(2b+1)E+(2b-a+\frac{1}{2})SS_x+(2b+1)FS $$
$$ W_2=W-W_1=FI+\frac{1}{2}S_x+ES+\alpha SS_y $$
$$ E = -\frac{i}{2\alpha} u_x,\,\,\,  F = \frac{i}{2}(\frac{(2a+1)u_{x}}{\alpha} -
2u_{y}) $$

Hence  we get the Lax refresentations of the M-VIII(26)
as $b = 0$ and for the Ishimori equation (29) as $a=b=-\frac{1}{2}$. The
M-IX equation(62)  admit the different type exact solutions
(solitons, lumps, vortex-like, dromion-like and so on).
As shown in [2] eqs. (62) have the following L-equivalent counterpart
$$ iq_t+M_1q+vq=0 \eqno(64a)$$
$$ ip_t-M_1p-vp=0 \eqno(64b)$$
$$ M_2v=M_1(pq) \eqno(64c)$$
where $p=E\bar q$. As well known these equations are too integrable [14]
and as in the previous case, equations (64) have the two integrable
reductions: equations(59) as $b=0$ and the Davey-Stewartson equation(61)
as $ a=b=-\frac{1}{2}. $ Note that the M-IX(62) and (64) equations
are g-equivalent to each other[15].

7) The M-XXII equation has the form[2]
$$ -iS_t=\frac{1}{2}([S,S_y]+2iuS)_x+\frac{i}{2}V_1S_x-2ia^2 S_y \eqno(65a) $$
$$ u_x=-\vec S(\vec S_x\wedge \vec S_y) \eqno(65b)$$
$$ V_{1x}=\frac{1}{4a^2}(\vec S^2_x)_y \eqno(65c)$$
The L-equivalent of these equations are given by[1]
$$ q_t=iq_{yx}-\frac{1}{2}[(V_1q)_x-qV_2-qrq_y]  \eqno(66a) $$
$$ r_t=-ir_{yx}-\frac{1}{2}[(V_1r)_x-qrr_y+rV_2] \eqno(66b) $$
$$ V_{1x}=(qr)_y  \eqno(66c) $$
$$ V_{2x}=r_{yx}q-rq_{yx} \eqno(66d)$$
where $r=E\bar q$. Both these set of equations are integrable and the corressponding
Lax representations were presented in[2].

8) The M-XXIX equation looks like[2]
$$ \vec S_{t}+\{\vec S_{xy}+V\vec S +
E\vec S_{x}\wedge (\vec S\wedge\vec S_{y})\}_{x}=0 \eqno(67a) $$
$$V_{x}=\frac{E}{2} (\vec S^{2}_{x})_{y}    \eqno(67b) $$
and the cooressponding L-equivalent counterpart of the M-XXXI
equation(67) has the form
$$ q_{t} +q_{xxy}-(qV_{1})_{x}-qV_{2}=0,    \eqno(68a) $$
$$ V_{1x}=2E(\bar q q)_{y}    \eqno(68b) $$
$$ V_{2x}=2E(\bar q q_{xy}-\bar q_{xy}q)    \eqno(68c) $$
which is the 2+1 dimensional complex mKdV. Note that
in the our case $E=\pm 1$.

9) The M-XXX equation has the following L-equivalent[2]
$$ q_t=iq_{yx}-\frac{1}{2}[(V_1q)_x-qV_2-Eqq_y] \eqno(69a) $$
$$ V_{1x}=Eq_y \eqno(69b) $$
$$ V_{2x}=-Eq_{yx} \eqno(69c) $$
$E=\pm 1$. Both these set of equations are integrable and the corressponding
Lax representations exist[2].
\\
\\
\\
\begin{center}
{\bf VI. ON THE INTEGRALS OF MOTION}
\end{center}

Of course the integrable nonlinear evolution equations
allows an infinite number of integrals of motion. The
above presented geometry allows construct the important integrals of motion.
So in 2+1 dimensions we have the following remarkable\\
{\bf  Theorema-1:}

The 2+1 dimensional nonlinear evolution  equations admit the following integrals of motion
$$
K_{1} = \int \kappa m_2 dxdy, \,\,\,\,\,K_{2} = \int \tau m_2 dxdy \eqno(70)
$$
 In particular for the 2+1 dimensional spin
systems this theorema we can reformulate in the following way\\
{\bf  Theorema-2:}

The 2+1 dimensional spin systems admit the following integrals of motion
$$ K_1 = \int \vec S \cdot (\vec S_{x} \wedge \vec S_{y} )dxdy \eqno(71a)$$
$$ K_2 = \int \frac {[\vec S \cdot (\vec S_{x}\wedge \vec S_{y})]
[\vec S \cdot (\vec S_{x} \wedge \vec S_{xx})]}
{\mid \vec S_{x}\mid^{\frac{5}{2}}}dxdy \eqno(71b)$$
Note that in the last case $G=\frac{1}{4\pi} K_1$ is the well known
topological charge.

If $\sigma \ne 0$ then  we have the following \\
{\bf  Theorema-3:}

The 2+1 dimensional nonlinear evolution  equations admit the following integrals of motion
$$
K_{3} = \int (\kappa m_2 +\sigma m_{3})dxdy \eqno(72a)
$$
$$
K_{4} =  \int (\tau m_2 + \sigma m_{1})dxdy \eqno(72b)
$$
$$
K_{5} =  \int (\tau m_3 -km_{1})dxdy \eqno(72c)
$$
The proves of these theoremes are given in[2].
\\
\\
\\
\begin{center}
{ \bf VII. ON THE LAX REPRESENTATIONS IN 2+1 DIMENSIONS}
\end{center}

It is of interest that the above considered formalism allows construct
the "Lax representations" the above presented nonlinear equations
[16].
\begin{center}
{\bf VIIa. The 2-dimensional case}
\end{center}
Let us introduce now the complex function

$$\phi_l={e_{2l}+ie_{3l}\over {1-e_{1l}}},\,\,\,\,e_{1l}^2+e_{2l}^2+e_{3l}^2=1,\,\,
\,\,\,l=1,2,3 \eqno(73) $$
The spatial and temporal evolution of this function can be written as a set
of the following equations:

$$\phi_{lx}={1\over 2} k \phi_l^2+
{1\over 2}k,\eqno(74a)$$

$$\phi_{ly}={1\over 2} m_3 \phi_l^2+
{1\over 2} m_3,\eqno(74b)$$

$$\phi_{lt}={1\over 2} \omega_3 \phi_l^2+
{1\over 2} \omega_3.\eqno(74c)$$

Further introducing the transformation

$$\phi_l={v_2\over{v_1}},\eqno(75)$$
eq.(74) can be written as

$$\pmatrix {
v_{1x} \cr
v_{2x}
}  = \pmatrix{
  0                  & {-k \over 2} \cr
{k \over 2} & 0
}
\pmatrix{
v_1 \cr
v_2
}, \eqno(76a) $$

$$\pmatrix {
v_{1y} \cr
v_{2y}
}  = \pmatrix{
0         & {{-1\over 2}m_3} \cr
{{1\over 2}m_3} & 0
}
\pmatrix{
v_1 \cr
v_2
}, \eqno(76b) $$

$$\pmatrix {
v_{1t} \cr
v_{2t}
}  = \pmatrix{
0  & {{-1\over 2}\omega_3} \cr
{{1\over 2}\omega_3} & 0
}
\pmatrix{
v_1 \cr
v_2
}. \eqno(76c) $$
These systems we can consider as the "Lax representation".
\\
\begin{center}
{\bf VIIb. The 3-dimensional case}
\end{center}

Introducing now the complex variable corresponding to an orthogonal rotation

$$\phi_l={e_{2l}+ie_{3l}\over {1-e_{1l}}},\,\,\,\,e_{1l}^2+e_{2l}^2+e_{3l}^2=1,\,\,
\,\,\,l=1,2,3 \eqno(77)$$
the spatial and temporal evolution of the trihedral  can
be rewritten as a set of the following three Riccati equations:

$$\phi_{lx}=-i\tau \phi_l+{1\over 2}\left[ k-i\sigma\right] \phi_l^2+
{1\over 2}\left[ k+\sigma\right],\eqno(78a)$$

$$\phi_{ly}=-im_1 \phi_l+{1\over 2}\left[ m_3+
im_2\right] \phi_l^2+
{1\over 2}\left[ m_3-im_2\right],\eqno(78b)$$

$$\phi_{lt}=-i\omega_1 \phi_l+
{1\over 2}\left[ \omega_3+i\omega_2\right] \phi_l^2+
{1\over 2}\left[ \omega_3-i\omega_2\right].\eqno(78c)$$

Further introducing the transformation
$$\phi_l={v_2\over{v_1}},\eqno(79)$$
eq.(78) can be written as a system of three coupled two component first order
equations,

$$\pmatrix {
v_{1x} \cr
v_{2x}
}  = \pmatrix{
{i\tau \over 2} & {-(k-\sigma) \over 2} \cr
{(k+\sigma) \over 2} & {-i\tau \over 2}
}
\pmatrix{
v_1 \cr
v_2
}, \eqno(80a) $$

$$\pmatrix {
v_{1y} \cr
v_{2y}
}  = \pmatrix{
{im_1\over 2} & {{-1\over 2}(m_3+im_2)} \cr
{{1\over 2}(m_3-im_2)} & {-im_1\over 2}
}
\pmatrix{
v_1 \cr
v_2
}, \eqno(80b) $$

$$\pmatrix {
v_{1t} \cr
v_{2t}
}  = \pmatrix{
{i\omega_1\over 2} & {{-1\over 2}(\omega_3+i\omega_2)} \cr
{{1\over 2}(\omega_3-i\omega_2)} & {-i\omega_1\over 2}
}
\pmatrix{
v_1 \cr
v_2
}. \eqno(80c) $$
These equations are the new form of the Lax representations of the
above considered 2+1 dimensional equations.
\\
\\
\\
\begin{center}
{ \bf VIII. A NON-ISOSPECTRAL PROBLEMS}
\end{center}

Unlike the 1+1 dimensions, where  $ \lambda_{t} = 0,$
as consequence the equation $ \lambda_{t} \ne const $ the 2+1
dimensional equations must be solve with help the non-isospectral
version of the inverse scattering transform(IST) method. In fact
that for example  for the M-III equations(54) the spectral
parameter $\lambda $ satifies the following equation[2]
$$\lambda_{t} = (2c\lambda^{2} + 2d\lambda)\lambda_{y} \eqno(81)$$
Of course that we can solve this eq. using the following Lax representation
$$ \Phi_x =[ic(\lambda^{2} - b^{2}) + id(\lambda-b)]\sigma_{3}\Phi \eqno(82a) $$
$$ \Phi_t =(2c\lambda^{2} + 2d\lambda)\Phi_y \eqno(82b) $$
For example we have the following two special solutions  of eq(81) as
$ c=d-1/2 =0$[2]
$$\lambda = \lambda_1= const  \eqno(83a)$$
$$\lambda=\lambda_2= {y+k+i\eta \over {b-t}}\eqno(83b)$$
where $b$, $k$ and $\eta$ are real constants. The corresponding
solutions of the soliton eqs.is we called the overlapping or
breaking solutions[17].

Note that unlike the 1+1 dimensions, where  $ \lambda_{t} = 0, $
in 2+1 dimensions  we have the following integral of motion for
the spectral parameter
$$ K = \int \lambda dy \eqno(84)$$
So $ K_{t} = 0$.
\\
\\
\\
\begin{center}
{ \bf IX. ON THE SOLUTIONS OF THE M-I EQUATION}
\end{center}

In this section we present the some exact solution of the M-I equation(47). Details
are given in[18].
The Hirota bilinear form of eq(47) are given by
$$ (iD_t-D_xD_y) (f^*\circ g)=0, \eqno(85a)$$
$$ (iD_t-D_xD_y) (f^*\circ f-g^*\circ g)=0, \eqno(85b)$$
$$D_x(f^*\circ f+g^*\circ g)=0,\eqno(85c)$$
while the potential $u$ takes the form
$$u(x,y,t)=-{iD_y(f^*\circ f+g^*\circ g)\over {f^*\circ f+g^*\circ g}},
\eqno(86)$$
where $g$ and $f$ are complex valued functions.
Here $D_{x}$  is the Hirota bilinear operator, defined by

$$D^{k}_{x} D^{m}_{y} D^{n}_{t} (f \circ g) = (\partial_x-\partial_
{x\prime })^k(\partial_y-\partial_{y\prime })^m (\partial_t-\partial_{t\prime })^n
f(x,y,t)g(x,y,t) \|_{x=x\prime, y=y\prime, t=t\prime} \eqno(87)$$
Using the above definition of the $D$-operator, we get from (31d) that

$$u_x = - 2i\left[ D_y (f \circ g) D_x(f^*\circ g^*)- c.c \right].\eqno(88a)$$
In terms of $g$ and $f$, the spin field takes the form

$$S^{+} = \frac{2f^{*}g}{\mid f\mid ^{2} + \mid g \mid ^{2}},\,\,\,\,\,
S_{3}=\frac{\mid f\mid^{2}-\mid g\mid^{2}}{\mid f\mid^{2}+\mid g\mid^{2}}.
\eqno(88b)$$
Eq.(85) represents the starting point to obtain interesting classes of
solutions for the spin system (47).The construction of the solutions is standard.
One expands the functions $g$ and $f$ as a series

$$g = \epsilon g_{1} + \epsilon^{3} g_{3} + \epsilon^{5}g_{5} +
\cdot \cdot \cdot \cdot \cdot, \eqno(89a)$$

$$f=1+\epsilon^2 f_2+\epsilon^4 f_4+\epsilon^6 f_6+ ..... . \eqno(89b)$$

Substituting these expansions into (85 a,b,c) and equating the coefficients
of $\epsilon $, one obtains the following system of equations from (85a):

$$\epsilon^1: ig_{1t}+g_{1xy}=0 ,\eqno(90a)$$

$$\epsilon^3: \left[ i\partial_t+\partial_x \partial_y\right] g_3=\left[iD_t
-D_xD_y\right] (f^*_2.g_1),\eqno(90b)$$
$$  \cdot                 \cdot $$
$$  \cdot                 \cdot $$
$$  \cdot                 \cdot $$

$$\epsilon^{2n+1}: \left[ i\partial_t+\partial_x \partial_y\right] g_{2n+1}
=\sum_{k+m=n} \left[ iD_t-D_xD_y\right] (f^*_{2k}.g_{2m+1}),\eqno(90c)$$
and

$$\epsilon^2: \left[ i\partial_t+\partial_x \partial_y\right] (f^*_2-f_2)=
\left[iD_t-D_xD_y\right] (g_1^*.g_1),\eqno(91a)$$

$$\epsilon^4: \left[ i\partial_t+\partial_x \partial_y\right] (f^*_4-f_4)=
\left[iD_t-D_xD_y\right] (g_1^*.g_3+g_3^*.g_1-f_2^*.f_2),\eqno(91b)$$
$$  \cdot                 \cdot $$
$$  \cdot                 \cdot $$
$$  \cdot                 \cdot $$

$$\epsilon^{2n}: \left[ i\partial_t+\partial_x \partial_y\right] (f^*_{2n}
-f_{2n})=(iD_t-D_xD_y)\left( \sum_{n_1+n_2=n-1}g_{2n_1+1}^*.g_{2n_(2+1)}\right)$$

$$-(iD_t-D_xD_y)\left(\sum_{m_1+m_2=n}f_{2m_1}^*.f_{2m_2}\right).\eqno(91c)$$
Further from (31c), we have the following:

$$\epsilon^2: \partial_x (f^*_2-f_2)=D_x(g_1^*.g_1),\eqno(92a)$$

$$\epsilon^4: \partial_x (f^*_4-f_4)=D_x(g_1^*.g_3+g_3^*.g_1+f_2^*.f_2),
\eqno(92b)$$
$$  \cdot                 \cdot $$
$$  \cdot                 \cdot $$
$$  \cdot                 \cdot $$

$$\epsilon^{2n}: \partial_x (f^*_{2n}-f_{2n})=
D_x\left[ \sum_{n_1+n_2=n-1}(g_{2n_1+1}^*.g_{2n_(2+1)}+
\sum_{n_1+n_2=n}f_{2n_1}^*.f_{2n_2})\right].\eqno(92c)$$
Solving recursively the above equations, we obtain many interesting classes of
solutions to eq.(47).

Using the results of the previous section, we are in a position to construct
many exact solutions such as solitons, domain walls, breaking solitons and
induced dromions of eq.(4). To obtain such solutions, we can use eqs.(85) as
the starting point.
\\
\\
{\bf A.   The 1 - line soliton solution}.

In order to construct exact 1-line soliton solutions (1-SS) of eq.(47),
we take the ansatz

$$ g_1= \exp {\chi_1},\,\,\,\,\chi_1 = a_1x + b_1y + c_1t + e_1.\eqno(93)$$
where $a_{1}, b_{1}, c_{1}$ and $e_{1}$ are complex constants. By sustituting
the above value of $g_1$ in eq.(90a), we get

$$c_1=ia_1b_1.$$
Hence $g_1$ reads as

$$g_1=\exp{\chi_1},\,\,\,\,\chi_1=a_1x+b_1y-ia_1b_1t+e_1.\eqno(94)$$
By picking up the appropriate equations from (90) to (92), we obtain the
expression for $f_2$ as

$$f_2=\exp{(\chi_1+\chi_1^*+\psi)},\eqno(95)$$
where
$$\exp{\psi}={-a_1^2\over {(a_1+a_1^*)^2}}.$$
By substituting the above values of $g_1$ and $f_2$ in eqs. (88b) and (88c),
we obtain the expressions for the spin components and for the potential.
\\
\\
{\bf B. The 2-soliton and N-soliton solutions}.

In this subsection we present the 2-SS of eq.(47).
To generate a $2$-line soliton solution (2-SS), we take

$$g_1=\exp{\chi_1}+\exp{\chi_2}.\eqno(96)$$
Substituting (96) in (88)-(90), after some calculation we obtain

$$f_2=N_{11}\exp {(\chi_1+\chi_1^*)}+N_{12}\exp {(\chi_1+\chi_2^*)}+N_{21}
\exp {(\chi_1^*+\chi_2)}+$$

$$N_{22}\exp {(\chi_2+\chi_2^*)},\eqno(97a)$$

$$g_3=L_{12}N_{11}N_{12} \exp {(\chi_1+\chi_1^*+\chi_2)}+L_{12}N_{22}N_{21}
\exp {(\chi_1+\chi_2+\chi_2^*)},\eqno(97b)$$

$$f_4=L_{12}L_{12}^*N_{11}N_{12}N_{21}N_{22} \exp {(\chi_1+\chi_1^*+\chi_2+
\chi_2^*)},\eqno(97c)$$
where

$$N_{rs}=-{k_r^2\over {(k_r+k_s^*)^2}},\,\,\,\,L_{rs}=-{(k_r-k_s)^2\over
{k_s^2}}.\eqno(98)$$
Inserting (97) into (88b) we get

$$S^{+}(x,y,t) =2{(1+f^*_2+f^*_4)(g_1+g_3)\over {\mid 1+f_2+f_4\mid^2+\mid
g_1+g_3\mid ^2}},\eqno(99a)$$

$$S_{3}(x,y,t)={{\mid 1+f_2+f_4\mid^2-\mid g_1+g_3\mid ^2} \over {\mid 1+
f_2+f_4\mid^2+\mid g_1+g_3\mid ^2}}, \eqno(99b) $$
and similarly the expression for the potential can also be obtained

By taking

$$g_N=\sum_{j=1}^N\exp{\chi_j}$$
and extending the above procedure, one can
obtain the N-SS also.
\\
\\
{\bf C.   The 1-curved soliton solution}.

In the one soliton solution, we can write the general form of $g$ as

$$g_1=\exp{\chi_1},\,\,\,\,\chi_1=a_1x+b_1(y,t)+e_1,\eqno(100)$$
where $b_1(y,t)$ is an arbitrary function of $y$ and $t$, satisfying the
relation

$$b_1(y,t)=b_1(y+ia_1t).\eqno(101)$$
Here also $a_1$ and $e_1$ are complex constants as stated earlier. Now
for the above general choice of the arbitrary function $b_1(y,t)$, the 1-SS
of eq.(47) takes the form

$$S_3(x,y,t)=1-{2 a_{1R}^2 \over {a_{1R}^2+a_{1I}^2}} sech^2{\chi_{1R}},
\eqno(102a)$$

$$S^+(x,y,t)={2a_{1R} \over {a_{1R}^2+a_{1I}^2}} \left[ia_{1I}-a_{1R}tanh
{\chi_{1R}}\right] sech {\chi_{1R}}, \eqno(102b)$$
while for the potential

$$u(x,y,t)={2a_{1R}\over {a_{1R}^2+a_{1I}^2}} \left(a_{1I}b_{1R}'-a_{1R}\eta
b_{1I}'\right) sech^2{\chi_{1R}} \eqno(102c)$$
in which the prime has been used to denote the differentiation with respect
to the real part of the argument. For fixed $(y,t)$, it follows from (102)
that  $\vec S \rightarrow (0,0,\pm 1)$ as $x \rightarrow \pm \infty$ and the
wavefront  itself is defined by the equation $\chi_{1R}=\eta x+m_{1R}(\rho)+
c_{1R}=0$. For other choices of $m_1$, we can obtain more interesting
solutions. \\
{\bf D.  The domain wall type solution}

The soliton solutions of (98) and (102) correspond to the boundary condition

$$\vec S(x,y,t) = (0. 0, 1),\,\, as\,\, x^{2} + y^{2} \rightarrow \pm \infty .
\eqno(103)$$
Another class of physically interesting solutions are the domain wall
type solutions, which have the asymptotic form

$$\vec S(x,y,t) = (0,0,\pm 1), as x^{2} +y^{2} \rightarrow \pm \infty \eqno(104)$$
In order to obtain domain wall solutions, we make
the choice

$$\omega(x,y,t) = g(x,y,t), \,\,\,\,\, f(x,y,t) = 1   \eqno(105)$$
Then, eq.(85) reduce to

$$ig_{t} + g_{xy} = 0,                 \eqno(106a)$$

$$g^{*}_{x} g_{y} + g^{*}_{y} g_{x} = 0,      \eqno(106b)$$

$$g^{*}_{x}g - g^{*}g_{x} = 0,                      \eqno(106c)$$
which is consistent with eq.(13).
Identically, we can use another way

$$\omega (x,y,t) = \frac{1}{f(x,y,t)},  \,\,\,\,g(x,y,t) = 1 \eqno(107)$$
Here also it follows from(85) that

$$if^{*}_t - f^{*}_{xy} = 0     \eqno(108a)$$

$$f^{*}_{x}f_{y} +  f_{x}f^{*}_{y} = 0 \eqno(108b)$$

$$f^{*}_{x}f -  f^*f_{x} = 0. \eqno(108c)$$
Comparing eqs. (106) and (108), we see that  eq.(47) is invariant under the
inversions, that is if $\omega (x,y,t)$ is the solution of eq.(47) then

$$ \omega ^{\prime }_(x,y,t) =\pm {1\over {\omega (x,y,t)}}, \eqno(109)$$
are also the solutions of eq.(47).

Now, we find the simplest non-trivial solutions for example of eq.(106). Let us
take the ansatz

$$g=\exp {(ax+iby-abt)} \eqno (110)$$
where $a$, $b$ are real constants. The components of the spin vector $\vec S$ are
given by

$$S^+(x,y,t)={\exp {iby}\over {cosh [a(x-bt-x_0)]}},\eqno(111a) $$

$$S_3(x,y,t)=-tanh [a(x-bt-x_0)].\eqno(111b) $$
We can also have a more general solution of the form

$$g=\exp {[ax+m(y,t)]},\eqno(112) $$
where $a$ is a real constant and $m(y,t)$ is an arbitrary function of $y$ and
$t$. From eq.(106a), it follows that

$$m=m(y+iat). \eqno(113) $$
Expressions for the spin components are then given by

$$S^+(x,y,t)={\exp [iIm(m(y,t))]\over {cosh [ax+Re (m(y,t))]}},\eqno(114a) $$

$$S_3(x,y,t)=-tanh [ax+Re (m(y,t))],\eqno(114b) $$
and the potential is

$$u(x,y,t)=-2Im(m(y,t)) \{ 1+\exp {[-2(ax+m_R)]}\}^{-1},\eqno(114c) $$
\\
\\
{\bf E.   The breaking soliton solution}.

We have already noted in Sec.II that for the present system (47), we have a
non-isospectral problem, as the spectral parameter $\lambda $ satisfies eq.(81).
The above presented solutions all correspond to the constant solution of
eq.(83a), namely $\lambda =\lambda_1=$ constant. One may consider other
interesting solutions of eq.(81). For example, one can have a special solution

$$\lambda=\lambda_1= \eta(y,t)+i \xi (y,t)={y+k+i\eta \over {b-t}},\eqno(115)$$
where $b$, $k$ and $\eta$ are real constants. Corresponding to this case, we may call
the solutions of eqs. (47) and (51) as breaking solitons[19].
Using the Hirota method, one can also construct the breaking 1-SS of eq.(47)
associated with (115). For this purpose, we take $g_1$ in the form

$$g=g_1= \exp{\chi },\,\,\,\, \chi = ax+m+c =\chi_R+i\chi_I ,\eqno(116) $$
where $a=a(y,t)$, $m=m(y,t)$ and $c=c(t)$ are functions to be determined.
Substituting (116) into the first of eq.(88), we get

$$ia_t+aa_y=0,\,\,\,\,im_t+am_y=0,\,\,\,iA_t+Aa_y=0, \eqno(117)$$
where $A=\exp(c)$. Particular solutions of eqs.(117) have the forms

$$ a=-i\lambda = {\eta -i(y+k) \over {b-t}},\,\,\,\, m=m{\left( y+k+i\eta
\over {{b-t}} \right) },\,\,\,\, A={A_0\over {b-t}} , \eqno(118)$$
where $\eta $, $k$, $b$ and $A_0$ are some constants. From eqs. (88)-(90),
we obtain

$$f_2 =B \exp {2\chi_R},\,\,\,\, B={\mid A_0 \mid^2 (y+k+i\eta )^2 \over
{4{\eta }^2(b-t)^2}} .\eqno(119) $$

Now, we can write the breaking 1-SS of eq.(47) (using equations (85b),
(116)-(119)),

$$ S^+(x,y,t) = {2\eta \exp {i(\chi_I+ \phi)}(y+k-i\eta ) \over {\left[ (y+k)^2
+\eta ^2\right] ^{3 \over 2}}} {\left[ (y+k)coshz -i\eta sinhz\right] \over
{cosh^2z}} ,\eqno(120a) $$

$$ S_3(x,y,t)= 1-{2\eta^2 \over {[(y+k)^2+\eta^2]}} sech^2z , \eqno(120b)$$
where $z={\eta \over(b-t)}x-{1\over2} ln[(y+\alpha )^2-{\eta }^2]+\psi $, \,\,\,
$ \psi = ln \mid  {A_0(y+k+i\eta ) \over {2\eta (b-t)}} \mid  $,\,\,\,
$ \chi_I= -{(y+k) \over {(b-t)}}x+m_I \left( {y+k+i\eta \over {b-t}} \right) $.
We see that the solution ( 120) corresponds to an algebraically decaying solution
for large $x$, $y$.
\\
\\
{\bf D.  Localized coherent structures (dromions)}.

Particularly, we present the dromion type localized solutions of eq.(4), the
so-called induced localized structures/or induced dromions[14] for the
potential $u(x,y,t)$.  This is possible by utilising the freedom in the choice
of the arbitrary functions $b_{1R}$ and $b_{1I}$ of $b_1$ in eqs.(101).
For example, if we make the ansatz

$$b_{1I}(\rho_R)= \kappa b_{1R}(\rho_R)= tanh(\rho_R), \eqno (121)$$

$$u=2 \eta (\xi -\eta \kappa) sech^2{\rho_R}sech[\eta x+ tanh{\rho_R}-
\eta x_0], \eqno (122)$$
where $\rho _R=y-a_{1I}t$. Similarly, the expressions for the spin can be
obtained from eqs.(98). The solution (122) for $u(x,y,t)$ decays exponentially
in all the directions, eventhough the spin $\vec S $ itself is not fully
localized. Analogously we can construct another type of ``induced dromion"
solution with the choice

$$b_{1I}= \kappa b_{1R} = \int {d \rho_R \over {(\rho_R+\rho_0)^2+1}}+b_0 ,
\eqno (123)  $$
where $\rho_0$ and $b_0$ are constants, so that

$$u(x,y,t)={2\eta (\xi - \kappa y)\over {(\rho_R+\rho_0)^2+1}} sech^2 \left [\eta x+
\int {d \rho_R \over {(\rho_R+\rho_0)^(2+1)}}- \eta x_0\right].\eqno (124)$$
\\
\\
\\
{\bf X. SOLUTIONS of (2+1) DIMENSIONAL NLSE}

In this section, we wish to consider briefly the corresponding solutions of
the equivalent generalized NLSE eq.(51). Already this equation has received
some attention in the literature. The following types of solutions are
available:
In a similar way, we can construct the N-breaking soliton solutions
of eq.(51). As an example, let us obtain the 1-breaking soliton solution
of eq.(51). The Hirota bilinear form of eq.(51) can be obtained by using the
transformation.

$$\psi={h\over \phi},\eqno(125)$$
as [19]

$$[iD_t+D_xD_y](h \circ \phi) = 0, \eqno(126a)$$

$$D_x^2(\phi\circ \phi)=2hh^*. \eqno(126b)$$
We look for the 1-breaking soliton solution in the following form:

$$h=\exp {\chi }, \eqno(127a)$$

$$\phi=1+\phi_2, \eqno(127b)$$
where$\chi =b(y,t)x+n(y,t)+c(t) $. Substituting (127) into (126), we get

$$ib_t+bb_y=0 , \eqno(128a)$$

$$in_t+bn_y=0 , \eqno(128b)$$

$$iB_t+Bb_y=0 , \eqno(128c)$$
and

$$\phi_2={\mid B\mid^2 \over {(b+b^*)^2}}\exp {\chi +\chi^*}=\exp {2(b_Rx+n_R
+\chi _0)}, \eqno (129) $$
where $\exp {2\chi_0}={\mid B\mid^2 \over {4b_R^2}}$, $B=\exp{c(t)}$ and
$b_R=b_R(t)=Re(b)$. Now, the formulae (125) provide us the 1-breaking soliton
solution of eq.(24)

$$\psi (x,y,t)={b_R(t)\exp {i\left[ b_I(y,t)x+n_I(y,t)+c_0\right] } \over
cosh \left[ b_Rx+n_R(y,t)+\chi_0\right] }, \eqno(130)$$
where $b(y,t)=b_R+ib_I$, $n(y,t)=n_R+in_I$ and $B(t)$ are the solutions of
eqs.(128). As for the case of eqs.(117), if we take the following particular
solutions of the system of eqs.(128);

$$\lambda ={y+k+i\eta \over {b-t}}, n= {y+k+i\eta \over {b-t}},
B={B_0\over (b-t)},\eqno(131)$$
then the 1-breaking soliton solution of eq.(51) takes the form

$$\psi (x,y,t)=-{n\over {b-t}}\exp{i\left[ {y+k\over {b-t}}x+n_I(y,t)+c_0\right] }
sech Z,\eqno(132)$$
where $Z={n\over {b-t}}x+n_R(y,t)+\chi_0$ and $c_0$, $\chi_0$ are constants.

Similarly, we obtain the breaking N-SS of (51). In this case we can take the
ansatz

$$g_{1} = \sum_{j = 1}^{N} \exp {\chi_{j}} \eqno(133)$$
with $\chi_{j} = b_{j}(y,t) x + n_{j}(y,t) + c_{j}(t)$. Inserting (133) into
(132) leads to

$$ib_{jt}+b_jb_{jy}=0,  \eqno(134a)$$

$$in_{jt}+b_jn_j=0,  \eqno(134b) $$

$$iB_{jt}+B_jB_{jy}=0,  \eqno(134c) $$
Proceeding as before, one can obtain breaking N-soliton solution(see, also,[20]).
\\
\\
\\
\begin{center}
{ \bf XI. ON THE SPIN-PHONON SYSTEMS}
\end{center}

Let us find the L-equivalent of the $M^{33}_{00}$ equation[2]
$$ \vec S_{t}=(\vec S\wedge \vec S_{x}+u\vec S\wedge \vec S_{x})_x \eqno(135a) $$
$$ u_t +u_{x} +\lambda (\vec S^{2}_{x})_{x}=0 \eqno(135b) $$
Let $ q=\tau-ik, \vec e_2\equiv\vec S, \sigma=\frac{1}{2\lambda}u-\frac
{1}{4\lambda}u^2. $ Then the L-equivalent of the equations (135) has the form
$$ iq_t+q_{xx}+(uq)_{xx}-\frac{i}{\lambda}[u-\frac{u^3}{2}+\frac{u^2}{2}]
q_x-\frac{1}{\lambda}[\frac{3i}{2}uu_x+(\frac{1}{4 \lambda}-4)u^2+
\frac{i}{2}u_x+\frac{u}{\lambda}+
\frac{u^3}{4\lambda}-\lambda |q|^2]q=0, \eqno(136a) $$
$$ u_t+u_x+\lambda |q|^2_x=0. \eqno(136b) $$

Similarly we can construct the L-equivalents of the other
spin-phonon systems[2].
\\
\\
\\
\begin{center}
{ \bf XII. CONCLUSION}
\end{center}

In this paper using the C-approach[2] we have presented the L-equivalent
soliton equations of the Ishimori and some Myrzakulov equations. Finally
we note that using the C-approach we can see to the older problems from
the new point of view. For example, the isotropic Landau-Lifshitz equation
$$ \vec S_{t}=\vec S\wedge \vec S_{xx} \eqno(137)$$
and the NLSE
$$ iq_t+q_{xx}+2Eq^{2}\bar q = 0 \eqno(138) $$
are L-equivalence to each other[2]. In the our case $E = \pm 1$. At the same time
the 1+1 dimensional M-IV and M-XXI equations have the following L-eqivalents:\\
the 1+1 dimensional mKdV
$$ q_t+q_{xxx}+6Eq^{2}q_{x} = 0 \eqno(139) $$
the 1+1 dimensional KdV
$$ q_t+q_{xxx}+6Eqq_{x} = 0 \eqno(140) $$
respectively.
\\
\\
\\
\begin{center}
{ \bf  ACKNOWLEDGMENTS}
\end{center}

I would like to thank Prof. M.Lakshmanan for hospitality during my visit
to  Bharathidasan University and for useful discussions.
The author are grateful for helpful conversations with Prof. J.Zagrodsinsky,
Dr. Radha Balakrishnan and Dr. M.Daniel.

\end{document}